\documentclass[fleqn]{mnras} 
\def\spose#1{\hbox to 0pt{#1\hss}}
\def\lta{\mathrel{\spose{\lower 3pt\hbox{$\mathchar"218$}}
     \raise 2.0pt\hbox{$\mathchar"13C$}}}
\def\gta{\mathrel{\spose{\lower 3pt\hbox{$\mathchar"218$}}
     \raise 2.0pt\hbox{$\mathchar"13E$}}}
 
\def\1p{\phantom{0}}
\def\2p{\phantom{00}}
\def\3p{\phantom{000}}
\def\4p{\phantom{0000}}

\usepackage{graphics,epsfig}
\usepackage{graphicx}

\title[The weak solar cycle 24]
{Carrington cycle 24: The solar chromospheric emission in a historical
and stellar perspective\thanks{based on TIGRE spectra, the NASA OMMNI data 
base, SORCE data provided by the LASP of the University of 
Colorado and SOLIS data obtained by the NSO}}

\author[K.-P. Schr\"oder, M. Mittag, J.H.M.M. Schmitt, D. Jack, 
A. Hempelmann, J.N. Gonz\'alez-P\'erez]
{K.-P. Schr\"oder$^{1}$\thanks{email: kps@astro.ugto.mx}, M. Mittag$^{2}$,
J.H.M.M. Schmitt$^{2}$, D. Jack$^{1}$, A. Hempelmann$^{2}$, \newauthor
J. N. Gonz\'alez-P\'erez$^{2}$  \\
$^{1}$Depto. Astronomia, Universidad de Guanajuato, 
A.P. 144, Guanajuato, GTO, C.P. 36000, Mexico 
\\
$^{2}$Hamburger Sternwarte, Universit\"at Hamburg, Gojenbergsweg 112, 
D-21029 Hamburg, Germany   }

\begin{document}

\label{firstpage}
\pagerange{\pageref{firstpage}--\pageref{lastpage}} 

\maketitle

\begin{abstract}
We present the solar S-index record of cycle 24, obtained by the TIGRE 
robotic telescope facility and its high-resolution spectrograph 
HEROS (R$\approx$20,000), which
measures the solar chromospheric Ca II H\&K line emission by
using moonlight. Our calibration process uses the same set of standard stars 
as introduced by the Mt.~Wilson team, thus giving us a direct comparison 
with their huge body of observations taken between 1966 and 1992, 
as well as with other cool stars. 
Carrington cycle 24 activity started from the unusually deep and long 
minimum 2008/09,
with an S-index average of only 0.154, by 0.015 deeper than the one of 
1986 ($<$S$>$=0.169). 
In this respect, the chromospheric radiative losses
differ remarkably from the variation of the coronal radio flux 
F10.7cm and the sunspotnumbers. In addition, 
the cycle 24 S-amplitude remained small, 
0.022 (cycles 21 and 22 averaged: 0.024), and so resulted a very low 2014 
maximum of $<$S$>$=0.176  (cycles 21 and 22 averaged: 0.193).  
We argue that this find is significant, since the Ca~II~H\&K 
line emission is a good proxy for the solar far-UV flux, which
plays an important role in the heating of the Earth's stratosphere,
and we further argue that the  solar far-UV flux changes changes
with solar activity much more strongly than the total solar output. 
\end{abstract}


\section{Introduction}

The period between approximately 1650 and 1715 has long been known for 
its paucity and/or absence of sunspots, first described by Sp\"orer (1890)
and Maunder (1894). For a long time, it had been thought that this apparent 
absence of sunspots was due to a lack of suitable observations, 
until Eddy (1976) convincingly demonstrated that this apparent absence 
of sunspots is really an actual absence of sunspots and not merely 
an absence of observations; Eddy (1976) also introduced the name 
``Maunder Minimum'' for this period of time. Ever since, the question 
has been around of what made the Sun enter and eventually leave 
this ''Maunder Minimum'' phase and what the properties of the Sun 
were at that time; unfortunately except for sunspot numbers 
no physical information is available from that period. 

The current Carrington Cycle 24 of the Sun follows the extraordinarily
long and deep minimum of 2008/09.
Schr\"oder et al. (2012) showed that there were ''zero-active''
minimum days, when the chromospheric emission was at its entirely
inactive, basal flux level. The Sun was entirely void 
of activity regions, not even showing plages or faculae, which are a
common sight on a ``normal minimum'', i.e., ''zero-sunspot'' day,
yet the random magnetic fine structure (unrelated to any magnetic 
activity region) is still present in 
those same days of zero-activity and not reduced at all, as seen in 
the respective SOHO images; as shown by V\"ogler and Sch\"ussler 
(2007), it is a simple by-product of convection (dubbed ``local dynamo'').

Various
solar activity data suggest that the maximum of Carrington cycle
24 occurred in the second half of 2014. The cycle was remarkably weak and 
even in the maximum year one day was recorded without any sunspots present 
on the solar surface (i.e., July 17, 2014).  By now, the Sun 
has already gone well beyond its peak activity in the current cycle and is 
approaching a new minimum. 
It is therefore appropriate to put the current solar 
cycle in its proper historical and stellar perspective.

\section{TIGRE-observations of solar chromospheric activity}

\subsection{Overview and context}

We use our TIGRE 1.2~m robotic telescope located in Guanajuato,
central Mexico, and its HEROS spectrograph (spectral resolution 
$\approx$ 20000, with
a spectral coverage of $\approx$~3800~{\AA} to 8800~{\AA}); this facility is 
described in detail by Schmitt et al. (2014). 

For a consistent treatment of the echelle order sampling,
assessment of their sensitivity functions, wavelength calibration
and straylight removal, a data-reduction pipeline is used. The
echelle order sampling parameters are recalibrated each time the 
spectrograph passes maintenance (typically once a year), while
flatfields, white-light and ThAr lamp emission line spectra are 
taken each night at least twice, before and after the observations.   
In addition, a spectrum of Vega or other suitable standard star is
taken to get a reliable, yet only relative flux-calibration.

Lunar light is used to obtain full-disk solar (i.e., Sun as a star) 
spectra typically twice a week whenever possible.
In this way we study the Sun with the same equipment as used for a set of 
standard stars (and program stars of our stellar chromospheric
activity survey), originally defined by O.C.~Wilson 
(see Baliunas et al. 1995, and Duncan et al. 
1991 and references therein).
Specifically, the surface area of the Moon sampled by the fiber of the
adapter unit channeling the light in the TIGRE spectrograph, has a size
of 3 arcseconds in diameter, and is thus only slightly larger than the average 
seeing disk of a star.  

Consequently, we can directly compare the integrated solar emission
to a huge body of stellar activity data gathered during the last 
five decades.

\subsection{Definition of the S-index}

The original Mt.~Wilson index is defined through the expression 
(Vaughan et al. 1978)
\begin{eqnarray}\label{sindex_mwo}
S_{\rm{MWO}} = \alpha \left( \frac{N_{H}+N_{K}}{N_{R}+N_{V}} \right),
\end{eqnarray}   
where $\alpha$ is an instrument-dependent calibration factor, used to put
data taken with different equipment on the same scale, $N_{H}$ and $N_{K}$
are the counts measured in a 1~{\AA} wide window, each centred on 
the Ca~II~H\&K line cores, and $N_{R}$ and $N_{V}$ are the counts in two 
20 {\AA} wide, ``quasi-continuum''  reference bands centred 
on 3901~{\AA} and 4001~{\AA}, conveniently encompassing the two Ca~II lines 
in between. In this way any variable atmospheric opacity and its
possibly variable slope in the near UV automatically cancel out. 
In the original four-photomultiplier set-up used at Mt.~Wilson,
the line-centred channels had an approximately triangular transmission 
profile, while we nowadays obtain digital records of spectra and
use a rectangular profile; the small differences between the two
are taken care of in the calibration process.

\subsection{Calibration of the TIGRE S-index}

For each instrumental setup, including our TIGRE facility, 
the calibration process has to be carried out continuously and must be
monitored carefully. For this purpose, TIGRE regularly observes a
set of stars, listed in Baliunas et al. (1995) and Hall et al. (2007) 
as a reference to all occurring levels of activity, and selected for 
their relative low degree of variation.

\begin{figure}
\centering
\begin{tabular}{c}
\epsfig{file=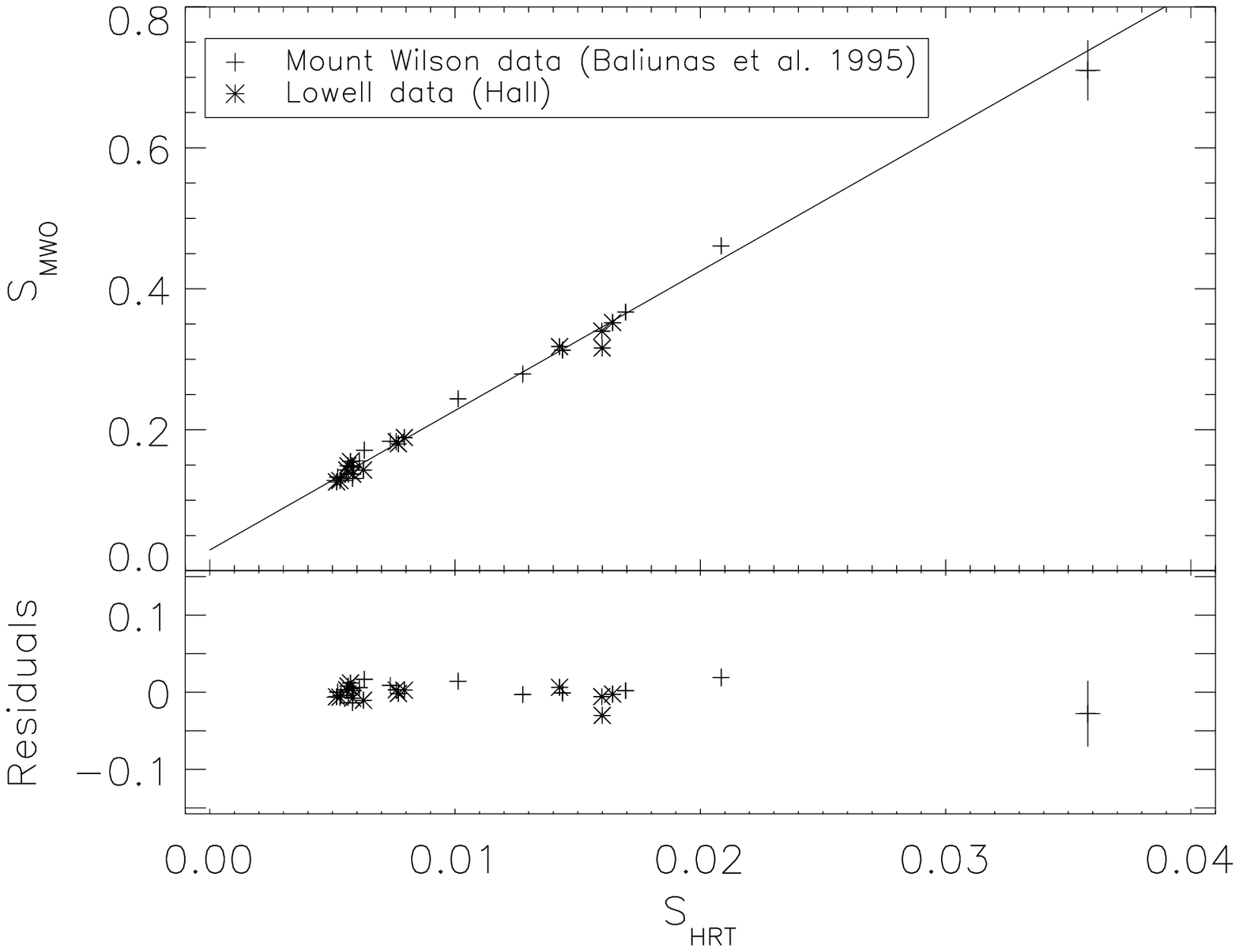,width=0.90\linewidth}
\end{tabular}
\caption{TIGRE S-index Calibration for the observations taken in Hamburg 
in 2008/09 (Mittag et al. 2011)}
\label{fig:calibration_2008/09}
\begin{tabular}{c}
\epsfig{file=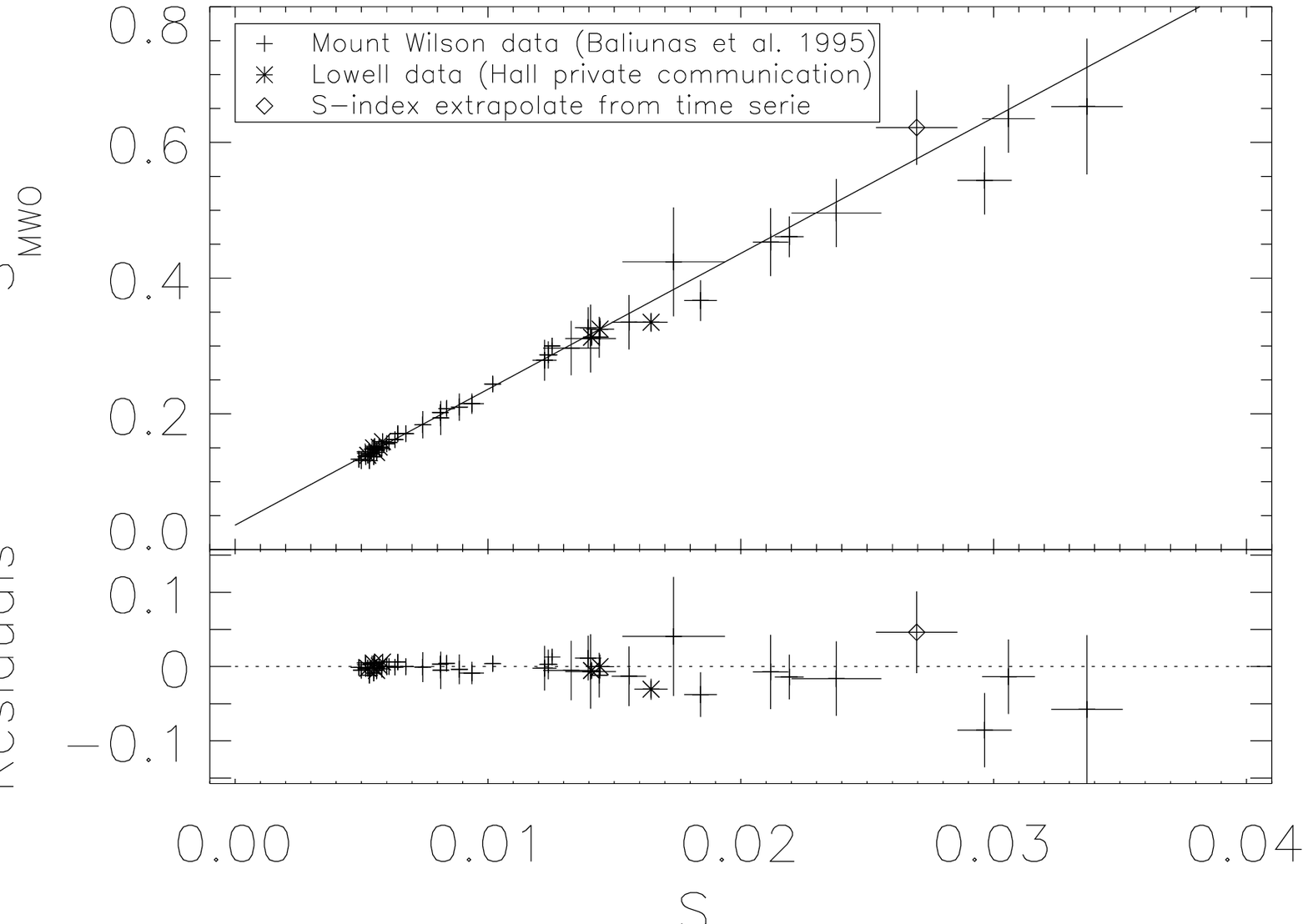,width=0.90\linewidth}
\end{tabular}
\caption{TIGRE S-index Calibration for the La Luz data 
taken since 2013 (Mittag et al. 2016)}
\label{fig:calibration_2013/14}
\end{figure}

Here, we show results from two different observation campaigns and
setups. The first campaign was undertaken in the season 2008/09 with
TIGRE still located at the Hamburg Observatory, then called Hamburg 
Robotic Telescope (HRT). 
The second campaign refers to the observations carried out at the 
La Luz Observatory since 2013. Therefore, we use two 
separate S-index calibrations 
to handle possible longterm instrumental drifts, 
such as residual under- or overcorrection for the echelle stray light 
in the line cores, which is accounted for with a zero-point
small offset in the linear regression of TIGRE's raw S-values over 
the respective original Mt. Wilson S-indices for our set of calibration 
stars. 
The calibration data for the period 2008/09 is shown in 
Fig.~\ref{fig:calibration_2008/09}, and in 
Fig.~\ref{fig:calibration_2013/14} for the data taken in
La Luz data since 2013. 
It is obvious from both figures that a linear
relation describes the transformation into the Mount Wilson scale very well;
specifically the calibration relation for 2008/09 data is given by 
(cf., Mittag et al. 2011): 

\begin{eqnarray}
S_{\rm{MWO}} = (0.029\pm0.004)+(19.8\pm0.6) S_{\rm{HRT}}
\end{eqnarray} 
and for the data taken since 2013 (cf., Mittag et al. 2016):
\begin{eqnarray}
S_{\rm{MWO}} = (0.036\pm0.003)+(20.0\pm0.4) S_{\rm{TIGRE}} .
\end{eqnarray} 

These relations are represented by a solid line 
in Fig.~\ref{fig:calibration_2008/09} 
and Fig.~\ref{fig:calibration_2013/14} together with the fit residuals.
We note that especially at low activity levels (critical for the Sun
in low activity states) the residuals are very small indeed, because
here the calibration stars show nearly no long-term variability.
Furthermore, the slope of the regression is the same in both
periods to within the errors, which demonstrates that the sample of 
calibrations stars as a whole does not significantly vary 
over longer timescales.

\begin{figure}
\centering
\epsfig{file=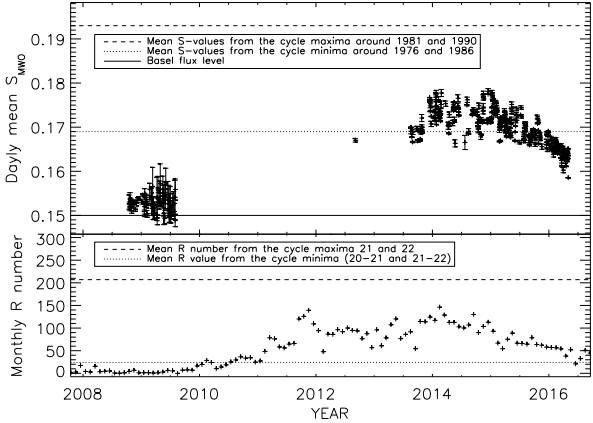,width=1.00\linewidth}
\caption{TIGRE's Mt.~Wilson-calibrated S-index values of the 2008/09 (last 
activity minimum) and 2013-16 period (error bars refer to measurement
uncertainties, only; calibration errors may reach 0.004), in 
perspective with previous average minimum and maximum activity (lines,
see section 4.1) and SIDC monthly sunspot numbers (below). 
The chromospheric emission of cycle 24 is much lower than 
in previous cycles.}
\label{fig:sindex_tigre}
\end{figure}

\subsection{TIGRE S-index of the Sun}

In Fig.~3 
we show the solar S-index measurements so far obtained with TIGRE.
While our TIGRE observations of the past solar minimum in 2008/09 were 
obtained from the preliminary telescope setup at Hamburg Observatory and 
published by Mittag et al. (2011) and Schr\"oder et al, (2012), 
the maximum of Carrington cycle 24 was covered from Aug.~2013 onwards, 
after the installation of TIGRE at the La Luz Observatory near Guanajuato. 
The best and simplest way to obtain the S-index of the Sun are
observations of lunar light, where the Moon is used as  
a ''mirror'' for the incident solar radiation.  Such observations
provide an unbiased measure of the light of the integral solar disk and
therefore such spectra record ``the Sun as a star'' with the
disk-integration coming for free.   

To test Olin Wilson's idea that the sunlight reflected by the Moon is a
very good proxy for direct sunlight and to check how well our reduction
pipeline deals with the echelle blaze functions, we
compare the respective near-UV spectral range with the
flux-calibrated, high-resolution KPNO FTS Solar Flux Atlas 
(see http://kurucz.harvard.edu/sun/fluxatlas/fluxatlastext.tab).
This atlas provides a very high SNR spectrum of integrated sunlight with
a spectral resolution of several hundred thousand.  Furthermore, the
FTS straylight and resulting solar spectrum dynamics are very well 
calibrated, and in the near UV the residual uncertainty is believed 
to be 0.2\%.

In Fig.~4 we compare the TIGRE lunar spectra
(black histograms) with the corresponding FTS spectra (red histograms)
for the four spectral ranges (V, K, H, R) relevant for the S-index
determination; note that the much higher spectral resolution FTS data
was ''blurred'' to the TIGRE resolution of $\sim$ 20000.  
Fig.~4 shows that the spectra agree very well,
and outside the cores of the Ca~II lines we find variations between our 
moon-light spectra and the FTS direct sun-light spectrum of under 1-2\%,
which appear on short scales and which we attribute to our broadening procedure.
In the cores of the Ca~II H\&K lines the FTS spectra are at a much higher level,
which is not surprising given they were taken at the time of solar maximum.

\begin{figure}
\centering
\begin{tabular}{c}
\epsfig{file=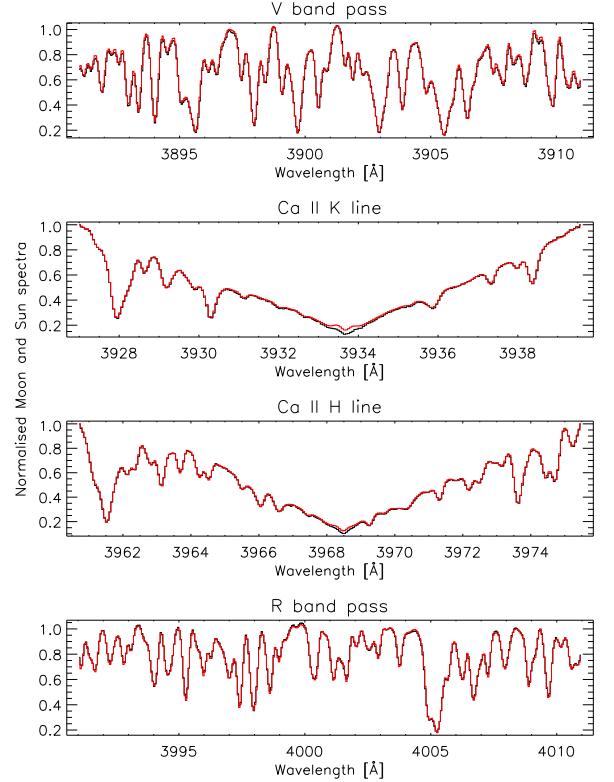,width=1.0\linewidth}
\end{tabular}
\caption{Comparison of a solar KPNO FTS spectrum (red; 
reduced to a spectral resolution of 20~000 for better comparability)
and a TIGRE moon-light spectrum
(black) in the spectral range relevant for the S-index measurements;
see text for details.}
\end{figure}


We therefore conclude that our TIGRE Moon spectra represent a very good 
approximation to the ''true'' solar spectrum in the Ca~II region,
and consequently our determinations of the solar S-index should have
a very small systematic error at worst. Since -- a benefit of our 
approach --  we use the same equipment and calibration for the Sun and 
for the stars, there is no instrumental or methodological bias at all and
consequently, the solar chromospheric activity recorded by means of the 
Mt.~Wilson S-index is {\it directly} comparable with the S-index records of 
stars. 

\begin{figure}
\centering
\begin{tabular}{c}
\epsfig{file=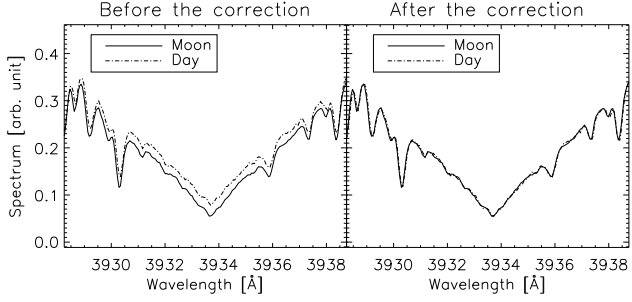,width=1.0\linewidth}
\end{tabular}
\caption{Moon and sky-scattered sunlight spectra before and 
after the scattering correction.}
\label{fig:sun_spec_day_corr}
\end{figure}

The method described above has been used for the data taken since 2013
at the La Luz site. 
However, in the observations taken in 2008/09 at Hamburg, only two direct
Moon spectra could be obtained. The majority of the Hamburg
observations uses sky-scattered sunlight. 
The physical processes involved in sunlight scattering include 
Rayleigh scattering, Mie scattering, as well as rotational Raman 
scattering off the N$_2$ and O$_2$ molecules in the Earth's atmosphere.
The latter process is particularly relevant for our Ca~II~H\&K data, since
it leads to the so-called ``Ring-effect'', first described by
Grainger \& Ring (1962).   Rotational Raman scattering is inelastic leading
to a a frequency redistribution of the scattered light, which leads 
to a fill-in of the stronger solar lines from the adjacent continuum.
Specifically, for the Ca~II H\&K lines, 
the Ring effect reaches well over 1\% of the solar continuum 
intensity and will therefore strongly affect the measurement of the S-index;
for a more recent, quantitative study of the Ring-effect we refer to
Sioris et al. (2002). 

The Ring-effect is naturally also present in the 2008/09 TIGRE data. 
In Fig.~\ref{fig:sun_spec_day_corr} we compare a solar spectrum 
(of the core of the Ca~II~K line) taken
from the Moon and form the day sky (left panel) and demonstrate that
our scattering corrections removes the Ring-effect (right panel).
We normalize the sky-scattered sunlight Ca~II~K line profile by the
deeper profile obtained from the direct sunlight obtained from the 
Moon spectra; our 
resultant correction of the sky-scattered sunlight spectra is of 
the same order as predicted by the modelling of Sioris et al. (2002).

\subsection{Further S-data-sets}

The gap in time between the original Mount Wilson S-measurements and our
TIGRE measurements
has been filled by several initiatives, most notably from Lowell Observatory
(J. Hall and colleagues) and the National Solar Observatory (NSO). 
However, the different calibration and instrumental issues  
still need a more detailed debate than the one we can give below. 

Using the late Mount Wilson, NSO/Sacramento Peak, Lowell and Lick 
S-values, Egeland et al. (2017) find that already cycle 23 was
quite weak, deriving maximum and minimum S-value of 
0.178 and 0.163 respectively. Inspecting their full data set of
S-index values, 
which is provided online and also covers cycle 24, we find good 
agreement with our S-index values for the time since 2013. However, 
for the past minimum, their S-index values are larger than ours, while 
their S-index values for the early 1990ies (cycle 22) are systematically 
lower than the measurements presented by Baliunas et al. (1995), 
to which we calibrated our S-index values shown in 
Fig.~\label{fig:sindex_tigre}. 

Egeland et al. (2017) argue that our TIGRE measurements taken in
2008/09 are affected by an incorrect correction for the Ring effect
described above. However, first, we argue that our corrections are
correct, and second, we point out the S-index values obtained from 
two direct moonlight spectra (without any Ring effect) in 2008, 
are also {\bf low}.  In addition, the S-index time series presented 
by Egeland et all (2017) does not appear to reflect, what is shown by 
all other solar activity 
indicators (see Section 2.6 and Section 3: SOLIS Ca II K 1-{\AA}, radio
flux, sunspot numbers): that the 2008/09 minimum is lower than the 
minima in 1986 and 1997.

\begin{figure}
\centering
\begin{tabular}{c}
\epsfig{file=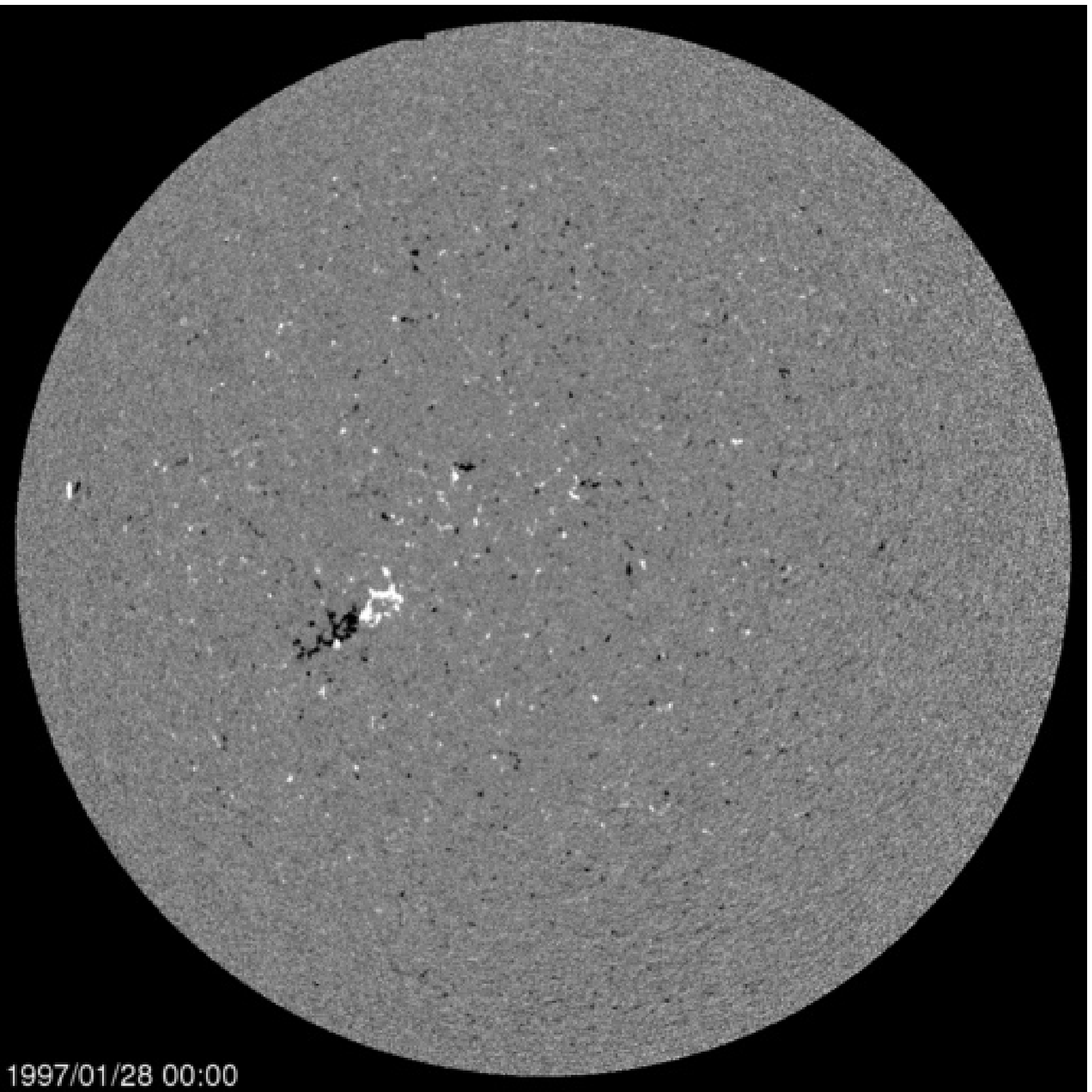,width=0.80\linewidth}
\end{tabular}
\caption{SOHO MDI magnetogram from  28 Jan 1997. (https://sohodata.nascom.nasa.gov/cgi-bin/data\_query)}
\label{fig:19970128_0000_mdimag_512}
\begin{tabular}{c}
\epsfig{file=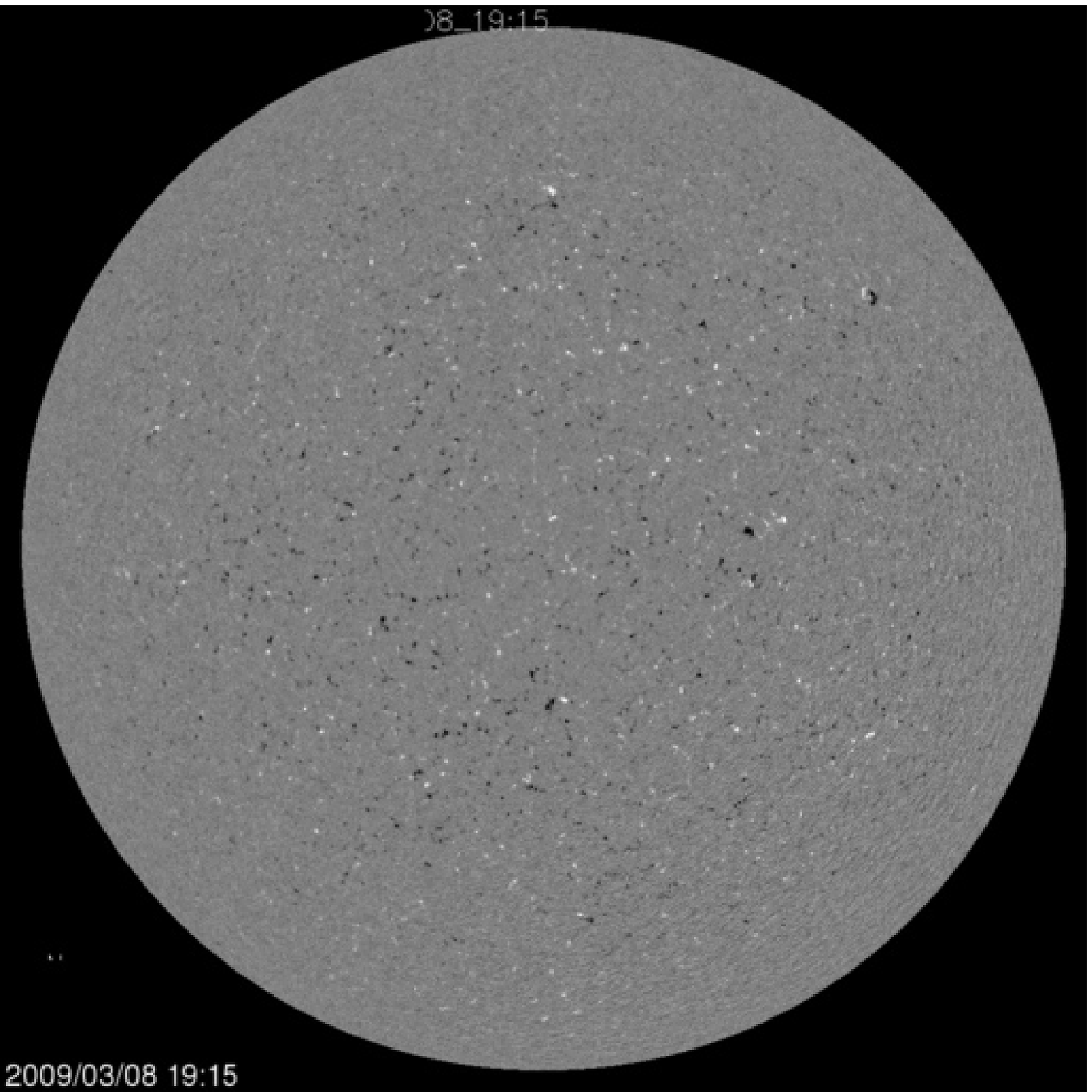,width=0.80\linewidth}
\end{tabular}
\caption{SOHO MDI magnetogram from 8 March 2009 (https://sohodata.nascom.nasa.gov/cgi-bin/data\_query)}
\label{fig:20090308_1915_mdimag_512}
\end{figure}

Further evidence is available from the fact that the Sun is the
only star with a directly observable surface. Egeland et al. (2017) 
inspected SOHO MDI magnetograms on two specific dates (Jan, 28$^{th}$
1997 and Oct, 9$^{th}$ 2009), since for both days they obtain relatively
low S-index values, of 0.1593 and 0.1596 respectively. 
The first value was measured with the original Mount Wilson 
HKP-2 instrument, while the second value was transformed from the 
NSO/Sacramento Peak Ca II K 1-{\AA} emissions index.   
While Egeland et al. (2017) state, that they find a low activity on
either day, a closer inspection of the MDI magnetogram from 
Jan, 28$^{th}$ 1997 (see Fig.~\ref{fig:19970128_0000_mdimag_512}) 
reveals the presence of a small region of chromospheric activity, while
by contrast, on Oct 9$^{th}$ 2009 the Sun was entirely void of any activity.
Yet the S-index quoted by Egeland et al (2017) for that day is not any
lower than the one measured by Mount Wilson for the slightly active
day Jan 28$^{th}$ 1997. 

We carried out the same exercise for Mar 8$^{th}$ 2009, since on
this day we obtained one of the two Moon light spectra from Hamburg
with a very low S-index of 0.152$\pm$0.006. 
In Fig.~\ref{fig:20090308_1915_mdimag_512} we shows the respective
SOHO MDI magnetogram taken on that day, and just as is the case for
Oct 9$^{th}$ 2009 there is absolutely no activity region visible
on the surface.  By this, admittedly limited but yet very representative 
probe, our S-index values appear to be consistent with the Mount Wilson 
measurements  during the past minimum period.

\subsection{Other solar data sets}

In the mid-seventies the National Solar Observatory (NSO) began taking regular 
observations of the disk-integrated solar Ca~II~K line at Sacramento Peak,
measuring in particular the so-called 1-{\AA} Ca~II~K emission index, 
defined as the equivalent width of a band with a width of 1-{\AA}
centered on the K-line core.
After October 1, 2015, the measurements were continued using the 
Integrated Sunlight Spectrometer (ISS) on the Synoptic Optical
Long-term Investigations 
of the Sun (SOLIS) facility. The ISS has actually been in operation since late 
2006 at NSO/Kitt Peak, so that there is an overlap of about eight years
between the two time series.

\begin{figure}
\centering
\begin{tabular}{c}
\epsfig{file=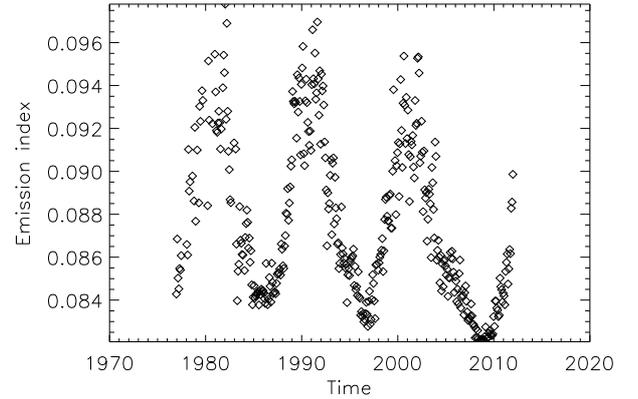,angle=0,width=0.99\linewidth}
\end{tabular}
\caption{Monthly medians of the SOLIS-calibrated Ca II K 1-{\AA}
index time series of the NSO (source: data made available 
by Bertello et al. 2017)}
\label{fig:SOLIS}
\end{figure}

Interestingly, the two time series differ and Bertello et al. (2017) provide
a detailed discussion of these discrepancies and develop a procedure to
unify the two time series.  Specifically, Bertello et al. (2017) correct the 
NSO/Sacramento Peak measurements  and empirically rescale them 
to the SOLIS data, which are not affected by 
sky-scattered light and the afore-mentioned Ring effect.  
In Fig.~\ref{fig:SOLIS} we
show the monthly medians of their re-calibrated data; obviously, the minimum in
NSO 1-{\AA} Ca~II~K emission index time series is clearly lower in
2008/09 than in the minimum in 1997, consistent  with the TIGRE S-index data.
Unfortunately, SOLIS is not able to measure the S-index directly and
cannot perform measurements of stars, therefore the stellar dimension
is missing from those measurements. 

\section{Solar activity during Carrington cycle 24} 

The last solar minimum and the ongoing Carrington cycle 24 are unusual not
only w.r.t. the solar UV emission, but also in
many respects, which we briefly demonstrate in this section.

\subsection{Radio emission and sun spot number}

\begin{figure}
\centering
\begin{tabular}{c}
\epsfig{file=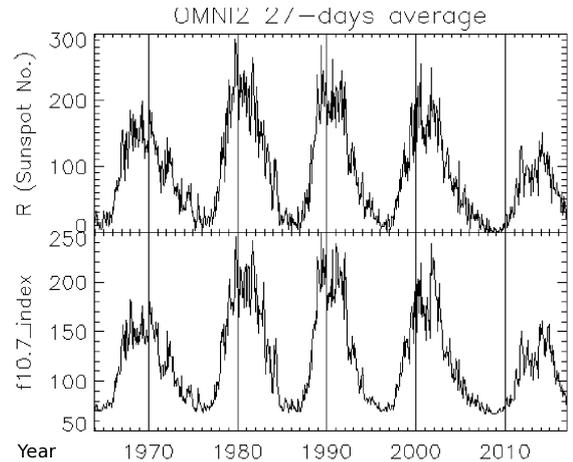,angle=0,width=0.90\linewidth}
\end{tabular}
\caption{27 day averages of: Upper panel: SIDC (new) sunspot 
number; lower panel: NRC F10.7cm coronal radio flux 
(source: OMNI/NASA).}
\label{fig:sol_radio}
\end{figure}

To provide a wider context we show in Fig.~\ref{fig:sol_radio} the monthly 
averaged sunspot numbers (upper panel, taken from the Belgian SIDC website) for the 
Carrington cycles 20 to 24 and  
values of the F10.7~cm radio flux (medium panel, NR Canada website). 
One recognizes, first, the very close correlation between these so different
activity indicators, and, second, that in the present cycle 24 
maximum, both these indicators reached only half of their levels
of the maxima of 1980 and 1991, or about their level of early 2003.  

\subsection{Total solar irradiance}

The total solar irradiance (TSI) has been continuously monitored by various
space-based instruments since 1978.   Since the zero-points
of the individual instruments do not fully agree, it is a non-trivial
task to combine the results of different missions into a definitive
record of the solar TSI.  BenMoussa et al. (2013) and 
Fr\"ohlich (2012) discuss the various
problems in extensive detail and provide a summary of the TSI
measurements  during the last three cycles (i.e., Carrington cycles 21
to 23). As discussed by Fr\"ohlich (2012),
even the TSI may have dropped during the
2008/09 minimum below the TSI values recorded during any of the 
previous minima, albeit only by a very small margin, probably  
smaller than $\Delta$TSI/TSI $\approx (10^{-4})$, but
long-term drifts and calibration differences between the 
different TSI missions prevent the derivation of more accurate numbers.

\subsection{Far UV-emission}

We next use the SOLSTICE data of the SORCE mission, 
provided by the Laboratory for Atmospheric and Space Physics (LASP) 
of the University of Colorado. These far UV 
spectral irradiance measures are of special interest, because
a ``haze'' of metal lines (Anderson \& Athay 1989) should reveal 
magnetic chromospheric heating by radiative losses, similar to 
the S-index, and like the latter, SOLSTICE data are star-calibrated,
with the consequence that they do {\it not} suffer 
from calibration lamp and/or electronics degradation problems (see
the discussion by Lockwood 2011 and references therein). 

\begin{figure}
\centering
\begin{tabular}{c}
\epsfig{file=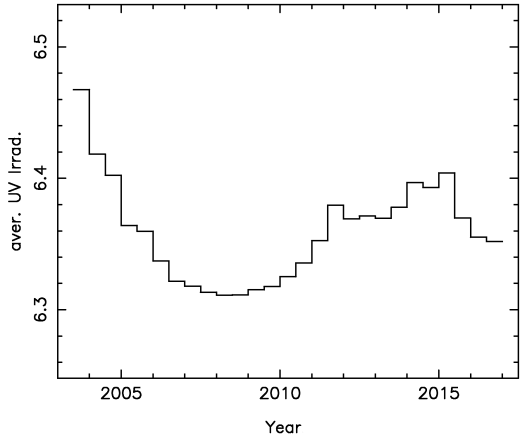,angle=-90,width=0.80\linewidth}
\end{tabular}
\caption{SOLSTICE/SORCE integrated spectral irradiance of the far-UV
200 to 280 nm window, relevant for stratospheric warming by
photodissociation of molecules, averaged over periods of half a year
between 2003 and 2016. The maximum of the present 24th cycle appears 
equally reduced as observed in TIGRE S-index time series calibrated to the
Mt.~Wilson scale. }
\label{fig:sun-uv}
\end{figure}

We integrate spectral irradiance over the wavelength range of 
200 to 280~nm, which by wavelength and intensity is the relevant 
spectral window for stratospheric warming by
photodissociation of molecules such as ozone and show 
(in Fig.~\ref{fig:sun-uv})
the averages obtained for periods of half a year, to minimize 
the impact of short-term fluctuations of the solar activity, covering 
the time-span of
2003 to 2016. Although this mission missed the maximum of cycle 23 by 
two years, it is evident from Fig.~\ref{fig:sun-uv} that the 
maximal far-UV irradiation in cycle 24 is, like the S-index, again
exceptionally low: 
With a total variation of only 1.4\%, the cycle 24  maximum level compares 
with the cycle 23 decline of late 2004. That is about 1\% lower than 
the far-UV irradiation of early 2003, with which time the cycle 24 
maxima of F10.7 and sunspot numbers compare.

\section{The historical solar and stellar perspective}

\subsection{The solar perspective}

\begin{figure}
\centering
\begin{tabular}{c}
\epsfig{file=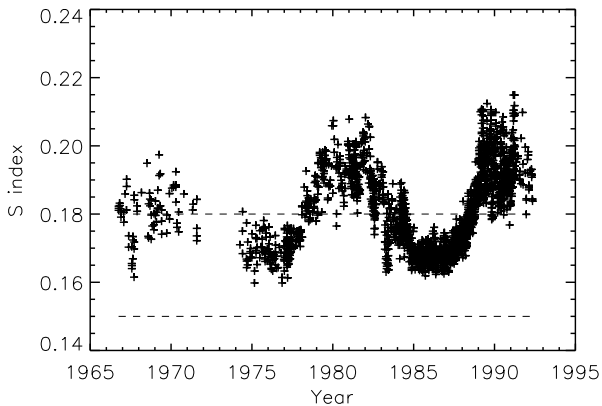,width=1.0\linewidth}
\end{tabular}
\caption{Historical solar Mt.~Wilson S-index values, covering the 
solar chromospheric
activity from the 1960ies to the 1990ies, according to Baliunas et al. 
1995. Horizontal lines mark extreme TIGRE S-index values of cycle 24. }
\label{fig:sindex_sun}
\end{figure}

The hitherto most complete description of the Mt.~Wilson H\&K 
monitoring project was presented by Baliunas et al. (1995).
Recently, these data have become publicly available, covering 
the period of 1966 to 1992 with rather unequal but sufficient 
sampling of the chromospheric activity maxima of 1980 and 1991 
(S-index of one-year-averages around maxima: 
0.193, extreme: 0.215), 
as well as of the minimum in 1986 (between cycles 21 
and 22, S-index one-year-average: 0.169, extreme: 0.162). 
Superimposed in Fig.~\ref{fig:sindex_sun} we plot (with dashed lines)
the extreme S-index values observed by TIGRE in 2008 and 2014, i.e.,
0.150 and 0.180 (cf., Fig.~\ref{fig:sindex_tigre}).  

We therefore conclude that the minimum in 2008 (average: 0.154) 
reached significantly 
below the minimum of 1986 (by 0.015 in one-year-averages), and the cycle 24 
maximum observed by TIGRE (S-index one-year-average of 0.176) is much lower 
(by 0.017) than the cycle 22 maximum; it is in fact closer to the
minimum of 1986. 

Hence, the change is not so much a smaller cycle 24 
S-index amplitude between the one-year-averages around maximum and
minimum (0.022, as compared to 0.024 in cycles 21 and 22). Rather,
chromospheric activity has now settled for a 0.015 deeper starting 
level. This is what makes the S-index underscore, on a relative scale, 
what the already low cycle 24 records of sunspot numbers and F10.7cm flux
let expect.

\subsection{The stellar perspective}

\begin{figure}
\centering
\begin{tabular}{c}
\epsfig{file=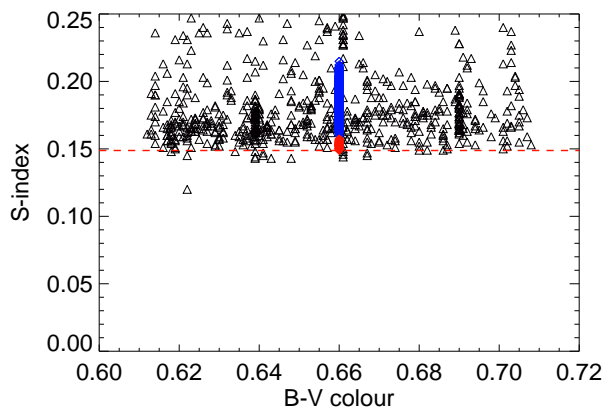,width=1.0\linewidth}
\end{tabular}
\caption{Stellar S-index values (triangles, taken from Mittag et al. 
2013) in the color range
0.61-0.71, together with the solar S-index values shown in 
Fig.~\ref{fig:sindex_sun} (blue filled symbols over B-V=0.66) 
and Fig.~\ref{fig:sindex_tigre} (red symbols); the red dashed line
marks the lowest TIGRE measured S-indices.
}
\label{fig:sindex_sun_stars}
\end{figure}

We next move on to the stars and compare -- 
in Fig.~\ref{fig:sindex_sun_stars} --  the solar S-index values
reported by Baliunas et al. (1995; blue data points) and our TIGRE S-indices 
(cf., Fig.~\ref{fig:sindex_tigre}, red data points) with a larger number
of S-index measurements of solar-like stars (taken from Mittag et al. 2013,
triangular data points).  Clearly,
the TIGRE values reproduce very 
well the lower cut-off (at around 0.15), observed for solar-like stars
already by Duncan et al. (1991).  We are therefore
confident that our TIGRE calibration is 
accurate, especially for stars of relatively low activity, and argue that 
during the extended solar minimum in 2008/09 the Sun was
actually close to the minimal S-index values observed for solar-like
stars and hence very close to the state it had during 
the ''Maunder Minimum'' period.

\section{Discussion and Conclusions}
The now outgoing Carrington cycle 24 is not only much weaker than its three
predecessors in terms of sunspot numbers and F10.7cm coronal radio flux
(see Fig.~\ref{fig:sol_radio}), the average S-index of
chromospheric emission was even more reduced. According to our 
TIGRE observations presented here, its amplitude between
one-year-averages of the S-index measurements was only 0.022 
(rather than 0.024, which was typical of cycles 21 and 22). Most remarkably,
however, the activity cycle started from a minimum deeper by 0.015,   
see Fig.~\ref{fig:sindex_sun}), meaning that the extraordinary
decrease of the S-index has a timescale longer than a single cycle.

The same behavior is seen in the averaged, integrated spectral 
far-UV flux in the wavelength range 200 to 280 nm 
(cf., Fig.~\ref{fig:sun-uv}), according to the 
star-calibrated SOLSTICE/SORCE data available for the timespan of 2003 
to 2016. Importantly, in both, the cycle 24 minimum (cf., Lockwood 2011) and
maximum (compare Fig. 3 with Fig. 2) far-UV irradiation is
lower by about 1\% against the projected cycle variation 
of the F10.7cm and sunspot number records. The cycle 24 amplitude 
of the far-UV averaged irradiation is 1.4\%. 
 
By simple comparison, we find that S-index and far-UV irradiation
cycle variations scale approximately to 0.016 in S-index being 
equivalent to 1\% in far-UV 
irradiation, making the former index a very good proxy for the latter
 -- not requiring space-based instrumentation and having a longer track 
record. We extrapolate, that in cycles 21 and 22 maxima, the average
far-UV irradiation must have been almost 3\% larger than in 2008, while
in the respective minima it was still 1\% larger.

Such a lasting deficiency of the solar UV irradiance implies a 
decline of the magnetic chromospheric heating which lasts beyond 
the present cycle, thus providing further evidence that the 
Sun is entering again a grand minimum phase.
In fact, such low activity could even been seen as   
the return to a long-term normal, since the two closest solar twins 
compare better, by their S-index values, with the low cycle 24 solar
chromospheric activity (see Mittag et al. 2016).   

In the context of possible TSI changes, we also 
analyzed how much  variance of the latter would be needed 
to produce a 1\% drop of the far-UV flux of the Sun, using PHOENIX model 
atmospheres (Hauschildt et al.~1999) with a very small difference in
effective temperature.  We find that, driven by the large UV opacities, a 
hypothetical decrease of the solar $T_{\rm eff}$ of as little as 
1.5 K over the decade or so in question would already account for such an 
additional 1\% decrease of the far-UV irradiation. However, this would also
imply a drop in TSI by about 0.1\%, probably more than observational 
evidence seems to suggest. Hence, we consider it more plausible that 
the decrease of the far-UV irradiation and Ca II line emission is
entirely due to a longterm reduced magnetic heating, which is
consistent with only a small extra decrease in the TSI during the past 
minimum.

The here presented evidence suggests that a phase of reduced solar activity
(or grand minimum) has an effect mainly on the solar far-UV emission, 
but much less so on the TSI. That would explain (see also the
discussion by Ermolli et al. 2013), why on the one 
hand there is only inconclusive evidence for a global cooling during the 
Maunder and Dalton minima but, on the other hand, there was a clear rise 
in the frequency of cold northern winters (as shown by Ineson et al 2012), 
which point to a then weaker jetstream and winter blocking situations. 

The hypothesis, albeit not yet reproduceable in global climate models, 
is that the above far-UV spectral range ($\approx$~200~nm to 280~nm) is
responsible for the photodissociation of molecules in the Earth's
stratosphere and constitutes its main radiative heating. 
Temperatures there are related to the strength and behavior of the 
jetstream (by climate sciences referred to as NAO, North Atlantic 
Oscillation). Consequently, low solar activity and
lower stratospheric temperatures seem to, statistically, favour a
weaker jetstream with wider oscillations (see Ineson et al.~2011 and 
references therein), subtly allowing for more pronounced blocking situations 
in the northern hemisphere. A rise of colder-than-average winters
during times of low solar activity is simply caused by the formation
of more large, persistent bubbles of arctic air -- without much affecting  
average world temperatures. 

In this context, we need consistent chromospheric emission records over long 
periods, in order to test refined climate models and to obtain a clearer 
correlation with climate records. Since direct, well-calibrated observations 
of the far-UV 
flux only exist for about 2 decades at best, the longer time-span available 
with the solar Mt.~Wilson-calibrated S-index records provide an important 
advantage. O.C. Wilson's star-calibrated method makes present 
and future records directly comparable with those 
obtained over half a century ago, and with TIGRE we will continue the 
spectroscopic monitoring of solar and stellar chromospheric activity.

\section*{Acknowledgments}

We acknowledge the very helpful travel support by 
bilateral projects CONACyT-DFG no. 192334 and PROALMEX 
(CONACyT-DAAD) No. 207772, as well as by CONACyT 
mobility grant No. 207662 (KPS), and by the 
DFG in several related projects. The HK\_Project\_v1995\_NSO data 
on the solis.nso.edu website derive from
the Mount Wilson Observatory HK Project, which ran from 1966 to 1992
and was supported by both 
public and private funds through the Carnegie Observatories, the 
Mount Wilson Institute, and the 
Harvard-Smithsonian Center for Astrophysics. It lived of the
dedicated work of O. Wilson, A. Vaughan, G. Preston, D. Duncan,
S. Baliunas, and many others. We also acknowledge discussions with
Drs. R.~Egeland and J. Hall, who provided their own solar S-index 
data set to us prior to publication, and Dr. M. Giampapa for drawing 
our attention to the work of Bertello and colleagues, clarifying the 
interpretation of the SOLIS K-index measurements.
Finally, and wish to thank our referee,
Dr. P. Judge, for his most helpful comments, 
which led to very substantial improvements.

{}

\label{lastpage}


\begin{thebibliography}{}

\bibitem{} Anderson L.S., Athay R.G. 1989, ApJ 346, 1010
\bibitem{} Baliunas S.L., Donahue, R.A., Soon, W.H., Horne J.H., Frazer J., 
           Woodard-Eklund L., Bradford M., and 20 coauthors  
           1995, ApJ, 438, 269
\bibitem{} Bertello, L., Marble, A.~R. and Pevtsov, A.~A., 2017, arXiv:1702.00838
\bibitem{} BenMoussa A., Gisso S., Sch\"uhle U., Del Zanna G., Auch{\`e}re F.,
   Mekaoui S., Jones A.R., Walton D., Eyles, C.J., Thuillier G. and 
   31 co-authors 2013, SoPh 286, 289    
\bibitem{} Duncan D.K., Vaughan A.H., Wilson O.C., Preston G.W., 
   Frazer J.L.H., Misch A., Mueller J., Soyumer D., Woodard L., 
   and 8 coauthors 1991, ApJS 76, 383
\bibitem{} Eddy J.A. 1976, Science 192, 1189
\bibitem{} Egeland R., Soon W., Baliunas S., Hall J.C., Pevtsov A.A., 
           Bertello L. 2016, arXiv:1611.04540v2 
\bibitem{} Ermolli I., Matthes K., Dudok de Wit T., Krivova N.A., 
           Tourpali K., Weber M., Unruh Y., and 8 coauthors 2013,
           Atmos.Chem.Phys. 13, 3945ff
\bibitem{} Fr{\"o}hlich C. 2012, Surveys in Geophysics 33, 453
\bibitem{} Grainger J.R., Ring J. 1962, Nature 193, 762 
\bibitem{} Hall J.C., Henry G.W., Lockwood G.W. 2007, AJ, 133, 2206
\bibitem{} Hauschildt P.H., Allard F. Baron E. 1999, ApJ 512, 377
\bibitem{} Ineson S., Scaiffe A.A., Knight J.R. et al. 2011, Nature
           Geoscience Letters, NGEO1282
\bibitem{} Lockwood M. 2013, J.Geophys.Res. 116, D16103
\bibitem[Maunder(1894)]{1894PASP....6..125M} Maunder, E.~W.\ 1894, PASP 
         6, 125 
\bibitem{} Mittag, M. and Hempelmann, A. and Gonz{\'a}lez-P{\'e}rez, J.~N. and 
	Schmitt, J.~H.~M.~M. and Hall, J.~C., 2011, ASPC, 448, 1187
\bibitem{} Mittag M., J.H.M.M. Schmitt, K.-P. Schr\"oder, 2013, A\&A  
         549, A117
\bibitem{} Mittag M.,  K.-P. Schr\"oder, A. Hempelmann, 
   J.N. Gonz{\'a}lez-P{\'e}rez, and J.H.M.M. Schmitt 2016, A\&A 591, A89
\bibitem{} Schr\"oder K.-P., Mittag M., P{\'e}rez Mart{\'i}nez M.I., 
           Cuntz M., Schmitt J.H.M.M. 2012, A\&A 540, A130
\bibitem{} K.-P. Schr\"oder, M. Mittag, A. Hempelmann, 
   J.N. Gonz{\'a}lez-P{\'e}rez, and J.H.M.M. Schmitt, 2013, A\&A 554, A50 
\bibitem{} J.H.M.M. Schmitt, K.-P. Schr\"oder, G. Rauw, A. Hempelmann,
   M. Mittag, J.N.~Gonz\'alez-P\'erez, S. Czesla, U. Wolter, D. Jack, 
   P. Eenens, M.A.~Trinidad 2014, Astron. Nachrichten, 335, 787--796
\bibitem{} Sioris C.E., Evans W.F.J., Gattinger R.L., McDade I.C.,
Degenstein D.A., Llewellyn E.J. 2002, Can.J.Phys. 80: 483--491
\bibitem[Spoerer(1890)]{1890AN....124..107S} Spoerer, F.~W.~G.\ 1890, 
     Astronomische Nachrichten, 124, 107  
\bibitem{} Vaughan, A.~H., Preston, G.~W., \& Wilson, O.~C. 1978, PASP, 90, 267
\bibitem{} V\"ogler A., Sch\"ussler M., 2007, A\&A, 465, L43
\bibitem{} Wright J.T. 2004, ApJ, 128, 1273
\end{thebibliography}
\end{document}